\documentclass[12pt, onecolumn, article, draft]{IEEEtran} 
\usepackage[dvips]{graphicx}
\usepackage{amsmath}
\usepackage{amssymb}
\usepackage{amsfonts}

\newcommand{\ignore}[1]{}

\begin{document}

\title{Comments on the Boundary of the Capacity Region of Multiaccess Fading Channels}

\author{
\authorblockN{Mohamed Shaqfeh and Norbert Goertz}\\
\authorblockA{Institute for Digital Communications\\
Joint Research Institute for Signal \& Image Processing\\ 
School of Engineering and Electronics\\
The University of Edinburgh\\
Mayfield Rd., Edinburgh EH9 3JL, Scotland, UK\\
Email:  \{M.Shaqfeh, Norbert.Goertz\}@ed.ac.uk
}}

\maketitle

\begin{abstract}
A modification is proposed for the formula known from the literature
that characterizes the boundary of the capacity region of Gaussian
multiaccess fading channels. The modified version takes into account
potentially negative arguments of the cumulated density function that
would affect the accuracy of the numerical capacity results.
\end{abstract}

\begin{keywords}
ergodic capacity region, multiaccess fading channel
\end{keywords}

\newpage

\section{Introduction}
The boundary of the capacity region of multiaccess (MAC) fading
channels was first characterized in \cite{TsHa:96} and discussed in
full detail in \cite{TsHa:98}. It is assumed that the
fading processes of all users are independent of each other, are
stationary and have continuous probability density functions, $f_i(h)$
$\forall i$, with $h\geq 0$ the random fading coefficient and $i$ the
user index; a total of $M$ users are assumed. The cumulated density
functions of the fading processes are denoted by $F_i(h) \doteq
\int_0^h f_i(h')dh'$. Note that, according to the standard fading
channel model with coherent detection, the support of the channel
coefficients does not contain negative numbers. The receiver noise is
assumed to be Gaussian with the variance $\sigma^2$.

\section{Boundary of the Capacity Region and Modification of the
  Standard Result}
\label{sec:ResCorrect}
It was shown in \cite[Theorem 3.16]{TsHa:98} that the boundary of the
capacity region of the Gaussian multiaccess channel is the closure of
the parametrically defined surface
\begin{equation}
\left\{ \mathbf{R}(\pmb{\mu}):\pmb{\mu}\in\Re^{M}_{+},
\sum_i\mu_i=1\right\}
\end{equation}
where for each $i=1,...,M$
\begin{equation}
\label{eq:rateold}
R_i(\pmb{\mu})=
\int\limits^{\infty}_{0}\frac{1}{2(\sigma^2+z)} \Bigg\{
\int\limits^{\infty}_{\frac{2\lambda_i(\sigma^2+z)}{\mu_i}}
f_i(h) \prod_{k\not=i}F_k\Big(
\underbrace{\frac{2\lambda_k
    h(\sigma^2+z)}{2\lambda_i(\sigma^2+z)+(\mu_k-\mu_i)h}}_{\doteq x}\Big) dh\Bigg\}  dz
\end{equation}
The vector $\pmb{\mu} \doteq \{ 0< \mu_i \leq 1: i =1,2,...,M \}$ is a
given ``rate award'' vector that is specified to pick a desired point
on the boundary of the capacity region. The vector
$\pmb{\lambda}\doteq \{\lambda_i \in \Re_+: i = 1,2,...,M\}$ is the solution of
the equations
\begin{equation}
\label{eq:powerold}
\int\limits^{\infty}_{0} \Bigg\{
\int\limits^{\infty}_{\frac{2\lambda_i(\sigma^2+z)}{\mu_i}}
\frac{1}{h}
f_i(h) \prod_{k\not=i}F_k\Big(
\underbrace{\frac{2\lambda_kh(\sigma^2+z)}{2\lambda_i(\sigma^2+z)+(\mu_k-\mu_i)h}
}_{\doteq x} \Big)dh
\Bigg\} dz =\bar{P}_i  \quad \text{for} \quad i =1,2,...,M\:,
\end{equation}
where $\bar{P}_i$ is the long-term average power constraint of user
$i$. The solution of (\ref{eq:powerold}) for the vector $\pmb{\lambda}$ is
unique, and an iterative numerical procedure is given in
\cite{TsHa:98} to find it.

As $0 < \mu_{i'} \leq 1$ $\forall i'$, the differences $\mu_k-\mu_i$
in (\ref{eq:rateold}) and (\ref{eq:powerold}) can have negative values
and, hence, the arguments of the cumulated density functions (CDFs)
can, depending on the channel coefficient $h$, also be negative. As
the fading coefficients can \textit{not} be negative, the CDF is
actually not defined for such values as they lie outside the support
of the random variable. Although it seems natural to assume the value
``zero'' in those cases, which might implicitly happen in a
implementation of (\ref{eq:rateold}) and (\ref{eq:powerold}), this
would lead to incorrect results as we show below.

To compensate for this problem, we propose to introduce a modified
argument in the cumulated density functions $F_k(x)$ in the
expressions in (\ref{eq:rateold}) and (\ref{eq:powerold}) as follows:
\begin{equation}
\label{eq:CumDensi}
    F_k(x) \xrightarrow[]{\text{replace}}  F_k([x]^\ast)  
\end{equation}
with 
\begin{equation}
x \doteq \frac{2\lambda_k
    h(\sigma^2+z)}{2\lambda_i(\sigma^2+z)+(\mu_k-\mu_i)h}
\end{equation}
and
\begin{equation}
\label{eq:Correct}
[x]^{\ast} \doteq \left\{
   \begin{array}{ll}
        x  & \text{if} \quad x \geq 0\\
        +\infty & \text{if} \quad  x<0
    \end{array} 
 \right. \: .
\end{equation}
For negative arguments, $x$, the function $[x]^{\ast}$ takes on the
value $+\infty$ which is inserted into a CDF in (\ref{eq:CumDensi}).
Hence the value of the CDF for $x<0$  is ``1'' and not
``0''. The justification is given in Section \ref{eq:Explanation}.

\section{Explanation}
\label{eq:Explanation}
There is no need to go through the whole derivation again to
characterize the capacity boundary surface. We start at the point
where we propose a modification, i.e., equation (18) on page 2804 of
\cite{TsHa:98}.  We wish to compute the rate
\begin{equation}
\label{eq:star_0}
R_i(\pmb \mu) = \int_0^\infty \frac{1}{2(\sigma^2 + z)} P(i,z) dz
\end{equation}
with 
\begin{equation}
\label{eq:start}
P(i,z) \doteq \Pr\Big(u_i(z) > u_j(z) \hspace{1em}\forall j 
\hspace{1em}\text{and} \hspace{1em}u_i(z)>0\Big)
\end{equation}
where the marginal utilities (``rate revenue minus power cost''
\cite[p.~2802]{TsHa:98}) are defined by
\begin{equation}
\label{eq:u}
u_i(z) \doteq \frac{\mu_i}{2\left(\sigma^2+z\right)}-\frac{\lambda_i}{h_i}
, \hspace{1em} z \geq 0  \;.
\end{equation}
To solve (\ref{eq:star_0}) (and also the corresponding problem in
\cite[equation (18)]{TsHa:98} for the
vector $\pmb \lambda$ to fulfil the average power constraint for the
user $i$) we need to evaluate the probability (\ref{eq:start}).

Firstly, it should be noted that the condition $u_i(z) > u_j(z) \:
\forall j$ in (\ref{eq:start}) (implicitly) excludes the case $j=i$
because otherwise $P(i,z)$ would be ``zero'' as, trivially, $P(u_i(z)
> u_i(z))=0$. Using (\ref{eq:u}) we can state the equivalence
\begin{equation}
\label{eq:Condition}
u_i(z)>0 \iff h_i>\frac{2\lambda_i(\sigma^2+z)}{\mu_i}>0 \:.
\end{equation}
Note that $\lambda_i>0 \:\forall i$, as $\lambda$ is a Lagrange
multiplier that introduces the ``power price'' (that can never be
negative) into the optimisation problem that must be solved to find
the capacity region \cite{TsHa:98}.

Using (\ref{eq:Condition}), the probability (\ref{eq:start}) can now be
written as
\begin{eqnarray}
P(i,z) & = & \Pr\Big(u_i(z)>0 \: \big| \: u_i(z) > u_j(z)\: \forall j
\Big) \cdot \Pr\big(u_i(z) > u_j(z)\: \forall j \big)\\
  & = &  \Pr\Big(h_i>\frac{2\lambda_i(\sigma^2+z)}{\mu_i} \: \big| \:
u_i(z) > u_j(z)\: \forall j \Big) \cdot \Pr\big(u_i(z) > u_j(z)\: \forall j
\big)\\
  & = & \int\limits^{\infty}_{\frac{2\lambda_i(\sigma^2+z)}{\mu_i}}
  f_i\big(h \: \big| \: u_i(z) > u_j(z)\: \forall j \big) dh \; \cdot  \;
\Pr\big(u_i(z) > u_j(z)\: \forall j \big) \\
& = &   \int\limits^{\infty}_{\frac{2\lambda_i(\sigma^2+z)}{\mu_i}}
  f_i\big(h, u_i(z) > u_j(z)\: \forall j \big) dh\\
& = &   \int\limits^{\infty}_{\frac{2\lambda_i(\sigma^2+z)}{\mu_i}}
  f_i(h) \cdot \Pr\Big(u_i(z) > u_j(z)\: \forall j  \: \big| \: h_i = h
  \Big) dh
\end{eqnarray}
%
\ignore{
The probability (\ref{eq:start}) can now be written as
\begin{eqnarray}
P(i,z) & = & \Pr\Big(u_i(z) > u_j(z)\forall j \: \big| \: u_i(z)>0\Big)
\cdot \Pr\big(u_i(z)>0 \big) \\
& = & \Pr\Big(u_i(z) > u_j(z)\forall j \: \big| \:
h_i>\frac{2\lambda_i(\sigma^2+z)}{\mu_i}\Big)
%
%
%
%
\int\limits^{\infty}_{\frac{2\lambda_i(\sigma^2+z)}{\mu_i}} f_i(h) dh \\
&  = &
\int\limits^{\infty}_{\frac{2\lambda_i(\sigma^2+z)}{\mu_i}}
f_i(h) \:
\Pr\Big(u_i(z) > u_j(z) \hspace{0.3em}\forall j 
\hspace{0.3em}\big| \hspace{0.2em}h_i=h \Big)dh  \; .
\end{eqnarray}
%
%
%
} 
%
Since the fading processes of the users are assumed to be independent,
we can write:
\begin{equation}
\label{eq:independent}
P(i,z) = \int^{\infty}_{\frac{2\lambda_i(\sigma^2+z)}{\mu_i}}
f_i(h) \cdot
\prod_{k\not=i}\Pr\left(u_i(z) > u_k(z) \hspace{0.3em}| 
\hspace{0.2em}h_i=h\right)dh \: .
\end{equation}

Now, we need to evaluate the probability
\begin{equation}
\label{eq:prob}
\Pr\left(u_i(z) > u_k(z) \hspace{0.3em}| 
\hspace{0.2em}h_i=h\right)
\end{equation}
We use (\ref{eq:u}) to rewrite the event $u_i(z)>u_k(z)$ and obtain
\begin{equation}
u_i(z)>u_k(z)   \iff   \frac{\mu_i}{a} - \frac{\lambda_i}{h_i}  > 
 \frac{\mu_k}{a} - \frac{\lambda_k}{h_k} 
\end{equation}
or, equivalently, 
 \begin{equation}
\label{eq:ineq_1}
 \frac{h_i (\mu_k -\mu_i) + \lambda_i a}{a \lambda_k  h_i} < \frac{1}{h_k}
\end{equation}
with the abbreviation $a \doteq 2(\sigma^2 + z)>0$ and
$\lambda_i>0$ $\forall i$  and $0<\mu_i \leq 1$ $\forall i$. As $\mu_k
-\mu_i$ can be negative, the left-hand side of (\ref{eq:ineq_1}) can
be negative so we have to differentiate between two cases:
\begin{eqnarray}
\label{eq:CaseA}
\text{Case A}:  & h_i (\mu_k -\mu_i) + \lambda_i a > 0  \iff 
   \big(\mu_k  \geq \mu_i\big) \quad \text{or} \quad   \Bigg( \mu_k<\mu_i \quad
   \text{and} \quad h_i < \frac{\lambda_i a}{\mu_i -  \mu_k} \Bigg)\\
\label{eq:CaseB}
\text{Case B}:  & \hspace{-19ex}  h_i (\mu_k -\mu_i) + \lambda_i a < 0   \iff
\mu_k<\mu_i \quad \text{and} \quad h_i >
\frac{\lambda_i a}{\mu_i -   \mu_k} 
\end{eqnarray}

\paragraph{Case A}

With $a=2(\sigma^2 + z)$ we obtain from (\ref{eq:prob}),
(\ref{eq:ineq_1}) and (\ref{eq:CaseA})
\begin{eqnarray}
\label{eq:prob_3_a}
\Pr\left(u_i(z) > u_k(z) \hspace{0.3em}|  \hspace{0.2em}h_i=h\right)
& = & \Pr\left( h_k <  \frac{2 \lambda_k  h_i (\sigma^2 +z) }{2  \lambda_i
  (\sigma^2 + z) +  (\mu_k -\mu_i)h_i} \Big| h_i = h \right)\\
& = & \Pr\left( h_k <  \frac{2 \lambda_k  h (\sigma^2 +z) }{2  \lambda_i
  (\sigma^2 + z) +  (\mu_k -\mu_i)h} \right)\\
\label{eq:prob_3_c}
& = &  F_k\left(\frac{2 \lambda_k  h (\sigma^2 +z) }{2  \lambda_i
  (\sigma^2 + z) +  (\mu_k -\mu_i)h} \right)
\end{eqnarray}
with $F_k(x) = \int_0^x f_k(h)dh$ the cumulated density function of
the channel coefficient $k$. The solution (\ref{eq:prob_3_c}) is the
one originally used in equations (\ref{eq:rateold}) and (\ref{eq:powerold})
that are taken from \cite{TsHa:98}.

\paragraph{Case B}

For a negative left-hand side in (\ref{eq:ineq_1}) we obtain 
\begin{eqnarray}
\label{eq:prob_4_a}
\Pr\left(u_i(z) > u_k(z) \hspace{0.3em}|  \hspace{0.2em}h_i=h\right)
& = & \Pr\left( h_k > B \right)  = 1
\end{eqnarray}
with 
\begin{equation}
 B \doteq \frac{2 \lambda_k  h (\sigma^2 +z) }{2  \lambda_i
  (\sigma^2 + z) +  (\mu_k -\mu_i) h} < 0  \;.
\end{equation}
As $h_k$ is a channel coefficient and non-negative by
definition, the probability (\ref{eq:prob_4_a}) is simply ``one''.

\paragraph{New formulation of the boundary of the capacity region}

In order to keep the structure of the original solution given in
\cite{TsHa:98} but with the correct evaluation of the probability in
both cases A and B, we write the probability 
\begin{equation}
\label{eq:Prob_Correct}
   \Pr\left(u_i(z) > u_k(z) \hspace{0.3em}|
   \hspace{0.2em}h_i=h\right) \: = \:  F_k\left(\Bigg[\frac{2 \lambda_k  h
   (\sigma^2 +z) }{2  \lambda_i  (\sigma^2 + z) +  (\mu_k -\mu_i)h}
   \Bigg]^*\right)
\end{equation}
with the function $[x]^*$ defined in (\ref{eq:Correct}). When we use 
(\ref{eq:Prob_Correct}) in (\ref{eq:independent}) and
(\ref{eq:star_0})  we obtain the corrected solution proposed in Section 
\ref{sec:ResCorrect}.


\section{Acknowledgements}
The work reported in this paper has formed part of the Core 4 Research
Program of the Virtual Centre of Excellence in Mobile and Personal
Communications, Mobile VCE, www.mobilevce.com, whose funding support,
including that of EPSRC, is gratefully acknowledged. Fully detailed
technical reports on this research are available to Industrial Members
of Mobile VCE.
The authors would also like to thank for the support from the Scottish
Funding Council for the Joint Research Institute with the Heriot-Watt
University which is a part of the Edinburgh Research Partnership.



\end{document}